\documentclass[preprint,12pt]{elsarticle}
\usepackage{amsmath}
\usepackage{amssymb}
\usepackage{graphicx}
\usepackage{bm}
\usepackage{url}
\usepackage{color}
\usepackage{ulem}
\usepackage{hyperref}
\journal{Computer Physics Communications}

\begin{document}
\begin{frontmatter}
\title{PLUMED 2: New feathers for an old bird}

\author[a]{Gareth A. Tribello \footnote{g.tribello@qub.ac.uk}}
\author[b]{Massimiliano Bonomi \footnote{mbonomi@salilab.org}}
\author[c]{Davide Branduardi \footnote{davide.branduardi@gmail.com}}
\author[d]{Carlo Camilloni \footnote{cc536@cam.ac.uk}}
\author[e]{Giovanni Bussi \footnote{bussi@sissa.it}}

\address[a]{Atomistic Simulation Centre, Queen's University Belfast, Belfast,
BT7 1NN}
\address[b]{Department of Bioengineering and Therapeutic Sciences, California Institute for Quantitative Biosciences, University of California, San Francisco, CA 94158.}
\address[c]{Theoretical Molecular Biophysics Group, Max Planck Institute for Biophysics, Max von-Laue strasse 3, 60438, Frankfurt am Main, Germany}
\address[d]{Department of Chemistry, University of Cambridge, Lensfield Road, Cambridge, CB2 1EW, United Kingdom}
\address[e]{International School for Advanced Studies (SISSA), Molecular and Statistical Biophysics, Trieste, TS, 34136, Italy}

\begin{abstract}
Enhancing sampling and analyzing simulations are central issues in molecular
simulation.  Recently, we introduced PLUMED, an open-source plug-in that
provides some of the most popular molecular dynamics (MD) codes with
implementations of a variety of different enhanced  sampling algorithms and
collective variables (CVs).  The rapid changes in this field, in particular new
directions in enhanced sampling and dimensionality reduction together with new
hardwares, require a code that is more flexible and more efficient.  We
therefore present PLUMED 2 here - a complete rewrite of the code in an
object-oriented programming language (C++). This new
version introduces greater flexibility and greater modularity, which 
both extends its core capabilities and
makes it far easier to add new methods and CVs.  It also has a simpler
interface with the MD engines and 
provides a single software library containing both tools and core facilities.  
Ultimately, the new code better serves the ever-growing community of users and
contributors in coping with the new challenges arising in the field.
\end{abstract}


\end{frontmatter}

\vspace{0.5cm}
{\bf PROGRAM SUMMARY}
\vspace{0.5cm}

\begin{small}
\noindent
{\em Manuscript Title: PLUMED 2: New feathers for an old bird}  \\
{\em Authors:} Gareth A Tribello, Massimiliano Bonomi, Davide Branduardi, Carlo Camilloni and Giovanni Bussi  \\
{\em Program Title:}  PLUMED 2                              \\
{\em Journal Reference:}                                      \\
{\em Catalogue identifier:}                                   \\
{\em Licensing provisions:}    Lesser GPL                  \\
{\em Distribution format:} tar.gz \\
{\em Programming language:}        ANSI-C++                           \\
{\em Computer:}      Any computer capable of running an executable produced by a C++ compiler         \\
{\em Operating system:} Linux operative system, Unix OS-es                     \\
{\em RAM:} Depends on the number of atoms, the method chosen and the collective variables used                           \\
{\em Number of processors used:}     1 or more \\
{\em Supplementary material:}   test suite, user and developer documentation, collection of patches, utilities                              \\
{\em Classification:} 3 Biology and Molecular Biology, 7.7 Other Condensed Matter inc. Simulation of Liquids and Solids, 23 Statistical Physics and Thermodynamics \\
{\em External routines/libraries:}        GNU libmatheval, lapack                          \\
{\em Nature of problem:}  calculation of free-energy surfaces for molecular systems of interest in biology, chemistry and materials science, 
on the fly and a-posteriori analysis of molecular dynamics trajectories using advanced collective variables
\\
{\em Solution method:} implementations of various collective variables and enhanced sampling techniques \\
{\em Unusual features:}
  PLUMED 2 can be used either as standalone program, {\it e.g.} for a-posteriori analysis of trajectories, or
  as a library embedded in a Molecular Dynamics code (such as GROMACS, NAMD, Quantum ESPRESSO, and LAMMPS). Interface with these software
  is provided in a patch form. Library is documented to ease its embedding into other software.
\\
  {\em Running time:} Depends on the number of atoms, the method chosen and the collective variables used \\
\end{small}

\section{Introduction}

Molecular simulations are now regularly used to understand natural phenomena in a wide range of 
subjects, spanning from biochemistry to solid state physics. These techniques are used to 
complement and interpret the increasingly large amounts of data coming from
experiments.  These tools are useful because two major technical advancements
make it so that it is now possible to extensively sample configuration space.
The first of these is the Moore's-law increase in computational power 
that has occurred during the last 50 years (see {\it e.g.}~\cite{de-shaw}). The
second, perhaps
more significant development, has been 
the creation of increasingly sophisticated algorithms
\cite{frenkel,tuckerman-book}. In the early days of simulations implementing
new 
algorithms was straightforward as every group had their own custom molecular dynamics (MD) code. 
Nowadays, however, there are a number of algorithmically-advanced simulation
suites \cite{dlpoly,plimpton1995fast,NAMD,gromacs,QE-2009,AMBER} that generally have a small number of 
technically-skilled developers and a much larger community of users.  These codes are complex and carefully optimized to 
work on modern, parallel computer hardware, which is  mandatory given the large computational facilities 
that are now available. However, this complexity makes experimenting and implementing new methods somewhat daunting.
Moreover, new methods end up being implemented in a specific suite, which rapidly becomes obsolete thus further 
slowing their dissemination.

For the above reasons we introduced the PLUMED plug-in a 
few years ago \cite{bono+09cpc}. This code was designed to ``plug-in" to MD 
codes such as
DL\_POLY\_CLASSIC \cite{dlpoly},
NAMD \cite{NAMD},
GROMACS \cite{gromacs} and
AMBER \cite{AMBER}
and to extend them by providing a 
single implementation for free-energy methods such as umbrella 
sampling \cite{torri-valle77jcp}, metadynamics \cite{laio-parr02pnas,WCMS:WCMS31} and steered 
MD \cite{grubmuller}. It was hoped that PLUMED would encourage
researchers in fields ranging from biophysics to solid-state physics to adopt
these methods. Furthermore, given that the interface with 
the complex MD code was looked after deep within PLUMED and that the coding style of PLUMED was 
both simple and flexible, it was hoped that developers would use the code to share their methods with 
the widest possible community.  PLUMED has been rather successful in both
respects and the plug-in model has been adopted by other researchers \cite{fiorin}.
To date there have been approximately 3200 downloads of PLUMED from our website,
more than 150 articles in which the original PLUMED article has been cited,
more than 200 users
who have subscribed to our mailing list, and 20 different super users who have contributed code fragments to our 
repository. PLUMED can now be interfaced with 10 
different MD codes and has even been used in ways we had not envisaged when the software was 
designed. Some particularly interesting developments being the use of 
PLUMED in at least three graphical tools that can be used to analyze trajectory data; namely, PLUMED 
GUI \cite{toniG}, METAGUI \cite{metagui} and GISMO \cite{sketchmap}.

The original PLUMED code was not designed to work with such a large variety of different MD codes or 
to grow this rapidly. As a result maintaining the interfaces between PLUMED and
the various MD codes  
has proved to be quite time-consuming.  Furthermore, the lack of a developer 
manual and programming guidelines has discouraged some contributors from sharing their code fragments and has contributed to
an untidy growth of the code.

Here we present PLUMED 2, a complete rewrite of the code aimed at addressing
the weaknesses in the original
design. 
In the new version, we have simplified considerably the interface 
between PLUMED and the MD codes so as to make this aspect of the code maintenance more 
straightforward.  PLUMED is now compiled separately as a software library and is thus completely 
independent from the MD codes. We have also moved to a 
modern, object-oriented programming language (C++) so to take advantage of inheritance and 
polymorphism. This has enabled us to use a plug-in architecture with a general
purpose core   
and functionality in separate modules.
It is thus easier to write bug-resilient code that can be worked on by 
multiple developers at the same time. Furthermore, developers 
can now easily modify 
the code and in principle even release independent extensions  thus bringing the 
PLUMED project closer to a community-based framework.  In addition, the object-oriented style allows us 
to write reusable objects whose functions are 
described in a developer manual that is generated from the code.
This makes it far easier to code the complex, multi-layered, nested collective variables (CVs)
that are increasingly being used to perform dimensionality reduction \cite{cecelia-review}. 
Finally, the new code structure allows one to use the same code for both on-the-fly biasing/analysis and post-processing thus minimizing redundancy. 

This paper is laid out as follows.  We first describe the theoretical 
background to PLUMED 2 and the various possibilities that it offers (section~\ref{sec:background}).  
We then describe how the code works (section~\ref{sec:code-operation}), the various standalone tools
that form part of PLUMED (section~\ref{sec:tools}) and the interface between a generic MD engine and PLUMED (section~\ref{sec:code-interface}).  We then provide a set of examples of varying complexity 
(section~\ref{sec:examples}) before finishing by showing how straightforward it 
is to implement new features and speculating a little on the various ways that this new code could be used 
in the future (section~\ref{sec:adding}). 

\section{Theoretical Background}
\label{sec:background}

In an MD simulation, the trajectories of a large number of atoms are calculated. The final result is thus a high 
dimensional description as to how the various atomic positions change as a function of time. 
It is very difficult to interpret the results of a simulation and to compare 
them with experimental data without further processing of the trajectory.  A
particularly useful way of processing the data 
is to calculate a histogram, $P(\mathbf{s})$, along a few selected CVs, $\mathbf{s}$, as from 
this one can calculate the free energy using:
\begin{equation}
F(\mathbf{s}) = -k_B T \log( P(\mathbf{s}) ),
\end{equation}
where $k_B$ is Boltzmann constant and $T$ is the temperature. This equation assumes 
that there are no high-energy barriers that prevent the system from visiting all 
the energetically-accessible portions of configuration space and 
that the trajectory is thus ergodic. Oftentimes this is not the case and
a commonly used technique to deal with this so-called time-scale problem is to add a bias along 
one or multiple CVs and to thereby force the system
to explore a wider range of CV values. In
these 
methods ({\it e.g.}~umbrella sampling, steered MD, metadynamics) the bias
$V(\mathbf{X},t)$ takes the form of an external 
potential, which may or may not be time dependent, but that is always a function
of some CVs, 
$\mathbf{s}(\mathbf{X}) =
\{ s_1(\mathbf{X}),s_2(\mathbf{X}),\dots,s_n(\mathbf{X})\}$:
\begin{equation}
V(\mathbf{X},t)=V(s_1(\mathbf{X}),s_2(\mathbf{X}),\dots,s_n(\mathbf{
X } ) , t) .
\end{equation}

A useful technique for providing greater flexibility in the code is to allow the user to construct new CVs 
as functions of other, simpler, CVs.  Even with this complexity though, it is still straightforward to 
differentiate this potential and thus obtain the force the bias applies to
the atoms using:  
                                                                               
 \begin{equation}
\frac{\partial V}{\partial x_m}=\sum_{i=1}^{n} \frac{\partial
V}{\partial s_i} \frac{\partial s_i}{\partial x_m}.
\end{equation}

This relation provides a powerful paradigm that can be used to design a flexible
plug-in.  The code can 
be divided into units that calculate the values and derivatives of the CVs, functions of CVs and biases in the equations 
above. The contribution the bias makes to the virial is calculated using a
similar procedure but involving derivatives of collective variables with
respect to the cell vectors.
Obviously, there are inter-dependencies between these quantities - it is not possible to calculate 
the value of $V(\mathbf{X},t)$ without first calculating the CVs, $\mathbf{s}(\mathbf{X})$. 
Consequently, the various quantities must be 
calculated in an order that starts with those that have an explicit dependence on the atomic positions. 

\section{PLUMED 2 Overview}

PLUMED 2 provides an executable and a C++ library. The most-basic
function of both these tools is the calculation of CVs from the
atomic positions. The benefit of having this functionality in a library as well
as in a standalone executable is that this allows one to calculate CVs during an
MD simulation.  This use will become more and 
more important in the near future as the available computing power is
increasing more rapidly than the 
volume of space that is available for storing trajectories.
 A second important function is PLUMED's ability to add additional
forces to 
the CVs as this is what allows us to implement free-energy methods such as
umbrella sampling, steered 
MD and metadynamics. PLUMED can also be used to perform other forms of analysis
of trajectory data. This can be done during post processing on the trajectory
file, much like a conventional analysis tool, or on-the-fly during the
simulation.
Other software is available for analyzing existing trajectories
\cite{gromacs,vmd,cpptraj,wordom1,wordom2,carma} or for biasing 
MD simulations \cite{fiorin,AMBER,desmond,orac,openmm,gromacs,NAMD}.  PLUMED 2,
however, is the only code we know of that allows users to do both sets of 
tasks with a single syntax.  This is important as a number of
recently-proposed algorithms 
\cite{gamus,recon} use the results from an analysis of a relatively short MD
simulation to refine the simulation bias.

PLUMED has now been completely redesigned to bring new features to both users and developers. 
Users will benefit from a new syntax for the input file that allows for much greater flexibility but which 
maintains strong similarities with the previous version. This flexibility can be exploited to create complex 
CVs without (or prior to) implementing them in C++.  This is possible because the functionalities that are 
already available can be combined together directly from the input file.  When this is done dependencies between 
quantities and chain rules for analytical derivatives are generated automatically. Users also benefit 
because PLUMED is now compiled independently from the underlying MD codes, which makes patching 
and including PLUMED in an MD engine considerably more straightforward. 
Lastly, virial contributions are calculated in PLUMED 2 so, unlike PLUMED
1, this code can be used to perform simulations with both constant
temperature and constant pressure.

The PLUMED 2 code now has a modular organization centered on a kernel of core functionalities. This 
allows developers to more easily contribute additional features, such as new CVs and free-energy 
methods, as they do not have to edit the core code. Dynamic polymorphism is used to add those features 
in a concurrent manner or even at run time. This flexibility can be achieved by
using C++ with minimal or 
no compromise on performance. In particular, many small utility classes are completely inlined, so as to 
obtain high level ({\it i.e.} easy to read) code running at maximum speed.

The PLUMED 2 executable can now also be used to run a few command-line tools.
These tools allow the user to both 
analyze existing trajectories and to run simple MD simulations.  In addition, there is an 
extensive set of regression tests. In these tests, PLUMED is used to analyze a set of trajectories and 
must reproduce exactly (within computational accuracy) a set of precomputed
results.  These tests can 
thus be used to ensure that new features are not introducing bugs into the software.  Finally, the 
PLUMED 2 library contains several small utilities, such as classes for treating periodic boundary 
conditions, functions for calculating root-mean-square deviation from reference
structures, and classes for editing and parsing
strings. Developers are provided with extensive documentation that describes all these classes as we believe 
that many of them could be used in applications outside of the PLUMED project.

In the following, we discuss in more detail some of the most important features
of PLUMED 2's design.  In particular, we will discuss the modularity of CVs and
biasing methods, both from the points of view of a user and a developer, the 
use of PLUMED 2 from the command line as a standalone tool and the interface
between PLUMED 
2 and a generic MD engine.

\subsection{How PLUMED 2 operates}
\label{sec:code-operation}

This section explains how PLUMED 2 operates when it is used to analyze or bias
MD simulations.  
PLUMED 2, much like the original PLUMED 1 package, takes a single dedicated input file in which 
each line instructs PLUMED to do something. An example input is shown below:

\begin{verbatim}
c1: COM ATOMS=1-10 
c2: COM ATOMS=30-40
d1: DISTANCE ATOMS=c1,c2 COMPONENTS 
a1: ANGLE ATOMS=14,15,16 
f1: COMBINE ARG=d1.x,d1.y,d1.z POWERS=2,2,2
t1: TORSION ATOMS=20,c1,c2,23
b1: METAD ARG=f1,a1 PACE=20 HEIGHT=0.5 SIGMA=0.05,0.1 
b2: UPPER_WALL ARG=d1.z AT=1.0 KAPPA=0.1
PRINT ARG=a1,t1,b1.bias,b2.bias FILE=colvar STRIDE=100 
\end{verbatim}

Each line in the input file above instructs PLUMED to create a new object, called an ``Action''. Every 
conceivable thing the user instructs PLUMED to do be it the calculation of a CV ({\it e.g.}~{\verb DISTANCE }, 
{\verb ANGLE } or {\verb TORSION } in the example) or a center of mass ({\verb COM }), 
the writing out of some data ({\it e.g.}~{\verb PRINT } in the 
above) or the calculation of a simulation bias ({\verb METAD } or {\verb UPPER_WALL }) can be cast as an Action object.  For the most part these Actions take in some input - usually either the positions of some of the atoms in 
the system ({\it e.g.}~for the first center of mass in the above the input is the positions of atoms 1 through 10)  or the 
instantaneous value of a CV - and use this data to calculate a new CV or bias potential. 
Obviously, Actions cannot be performed in an arbitrary order as Actions that take CVs as input clearly 
cannot be computed without first calculating the prerequisite CVs. Hence, PLUMED, while reading the 
input, ensures that the data required by each Action is either available without calculation from the 
trajectory or is part of the 
output from the Actions that precede it in the input file. The user should thus think of the PLUMED input file 
as a kind of primitive script that provides a set of instructions that will be executed during the simulation or during the analysis. 

Most of the collective variables and methods that were available in
 PLUMED 1 can be recoded in PLUMED 2 using either a single Action or using a
small number of
  Actions.  Furthermore, the PLUMED 1 variables  that have been recoded in
PLUMED 2 contain fewer 
  lines of executable code and many of them have a greater flexibility than
their PLUMED 1 counterparts. 
 At present we 
  have reimplemented the basic geometric quantities (distances, angles and torsional angles) as well as 
  quantities such as coordination numbers, minimum angles and alpha-beta similarities 
  \cite{pian-laio07jpcb} that are non-linear combinations of these simpler quantities.  We have also 
  implemented the all important root-mean-square deviation \cite{kearsley} and
have used these routines to 
  write Actions to calculate the path CVs \cite{bran+07jcp} and protein secondary structure variables 
  \cite{Pietrucci:2009wv} that were in PLUMED 1 and the generic property map that was 
  recently proposed by Spiwok and Kr{\'a}lov{\'a} \cite{IsoMetad}.   As well as these generic CVs 
  PLUMED 2 contains implementations of CVs that are used by specific communities.  For researchers
  examining polymers we have implemented the radius of gyration as well as inertia-tensors-based 
  CVs \cite{Vymetal:2011gv}.   Users can also calculate the total volume of the cell, the total potential
  energy of the system \cite{PhysRevLett.104.190601} or the dipole formed by a set of charged atoms.  Lastly
PLUMED 2 contains
  Actions for calculating the Debye-H\"uckel energy \cite{do+13jctc} and for
interfacing PLUMED with 
  the Almost library so that the CamShift Collective Variable 
\cite{Robustelli:2010dn,Camilloni:2012je} 
  can be calculated.

\begin{figure}
\centering
\includegraphics[height=8cm]{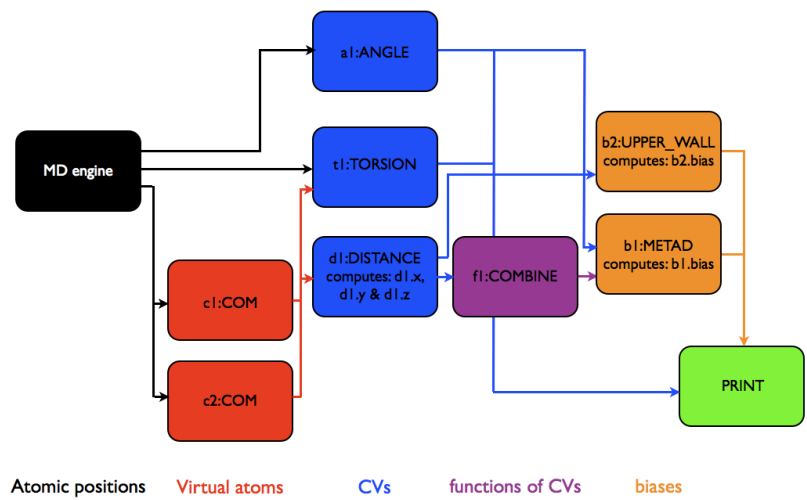}
\caption{A diagram that shows how data is passed between the Actions in a PLUMED job.}
\label{fig:code-passing}
\end{figure}

Each of the Actions defined in the PLUMED input is given a unique label by the user. This label can be 
used to retrieve the output from the Action so that it can be used in a later part of the calculation. As an 
example of how this works in practice each of the first two commands in the input file above instructs 
PLUMED to calculate the position of a center of mass. These two centers of masses are stored in 
containers labeled {\verb c1 } and {\verb c2 } and are used when PLUMED calculates the distance between center of 
mass {\verb c1 } and center of mass {\verb c2 } as part of the Action labeled {\verb d1 }.

 The first eight  Actions defined in the input above all do something similar - {\it i.e.} they all fill a container with the various 
quantities that are calculated during the Action's execution. When only a single quantity is calculated this 
quantity is referenced in the later parts of the input file using the Action's label. To reduce computational 
overhead some Actions calculate multiple quantities\footnote{This is
particularly useful for 
quantities such as path CVs, $s$ and $z$, which are just different non-linear combinations of 
some expensive to calculate base functions (for $s$ and $z$ a set of
mean-square deviations)}.  The values output by these quantities can be 
referenced using $\langle$label$\rangle$.$\langle$component$\rangle$. 
As an example, the keyword {\verb COMPONENTS } in the DISTANCE Action in
the above instructs PLUMED to store the $x$, $y$ and $z$ components of the
distances separately.  These separate values can then be referenced using
the labels  {\verb d1.x },  {\verb d1.y } and  {\verb d1.z } as
they are above in the {\verb COMBINE } Action with label {\verb f1 }.  The
names of the quantities that are calculated by any Action are provided in the
manual and in the output at run-time so that
the user can correctly refer to quantities when writing the input.
Figure~\ref{fig:code-passing} illustrates the manner in which data is passed between Actions more 
clearly by showing schematically what quantities are calculated by each of the Actions in the above input 
and how this data is passed about. 

\begin{figure}
\centering
\includegraphics[height=8cm]{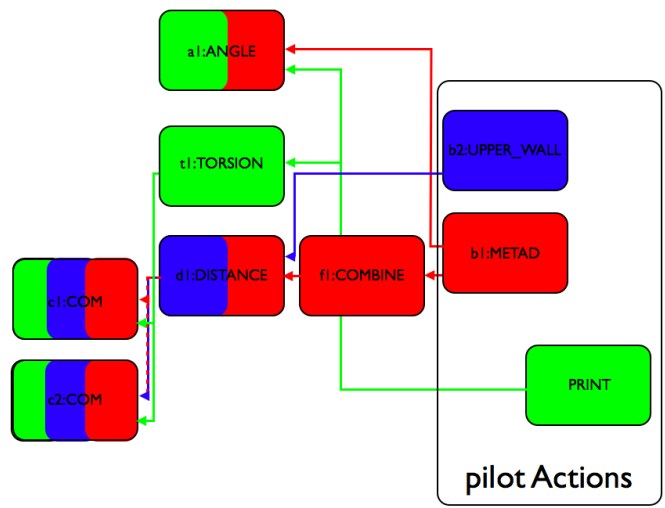}
\caption{Schematic representation of how the Actions executed during each step of a PLUMED calculation 
are controlled by a small number of pilot Actions. The user specifies the frequencies with which these 
pilots should be performed. At each step PLUMED works out which pilots are to be performed and then 
activates all the Actions which calculate the data that is required by the active pilots.}
\label{fig:code-depend}
\end{figure}

The passing of data between the Actions in the input file introduces a clear set of interdependencies 
between them. This is illustrated for the input defined above in figure~\ref{fig:code-depend}. As this figure 
shows, one cannot calculate the metadynamics bias without first calculating the function {\verb f1 } and the angle 
{\verb a1 }. Similarly one need only calculate the torsion angle labeled {\verb t1 } during those steps when 
the {\verb COLVAR } file is written. PLUMED thus uses these interdependencies to control when the
various Actions should be 
executed. The user provides instructions in the input as to the frequency with which certain, so called pilot 
Actions (in the example input above {\verb METAD }, {\verb UPPER_WALL } and {\verb PRINT } Actions) should be performed. 
Then, when PLUMED is called, it examines the list of pilot Actions, establishes which 
of them must be performed at the current time and activates the full set of Actions on which each of the active pilots 
depends. Once this process is completed PLUMED executes the set of activated Actions and calculates 
everything that is required at the current time step. This pre-screening of the
Actions  
saves computational effort by ensuring that expensive CVs are only calculated when they 
are needed.   It therefore considerably lowers the execution time for the calculation especially when 
expensive CVs are monitored only rarely.

\begin{figure}
\centering
\includegraphics[height=8cm]{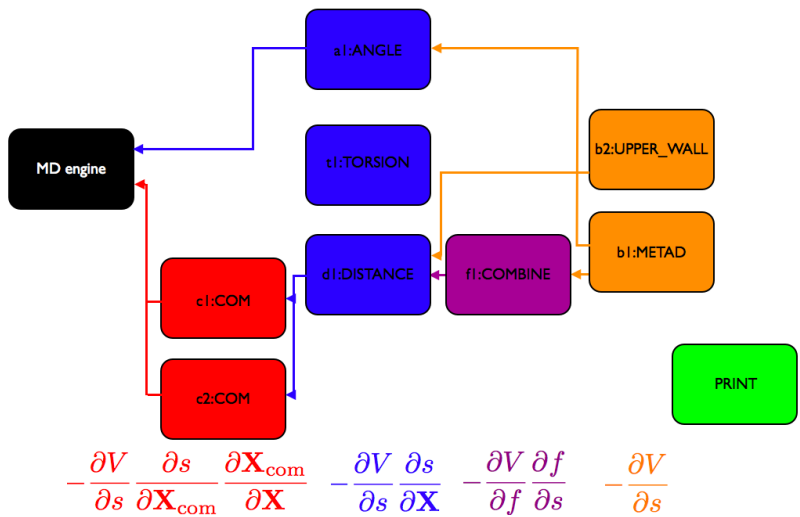}
\caption{Figure showing how the forces from the simulation bias are transferred onto the atoms 
using the chain rule.}
\label{fig:code-apply}
\end{figure}

When one is using PLUMED to perform analysis on an MD trajectory, the calculation is 
finished once the values for each of the Actions has been calculated. By contrast when one is using 
PLUMED to bias the dynamics a second step is required as the forces from the bias must be transferred 
onto the atoms. This sort of calculation is possible because during the initial calculation step the 
derivatives of the CVs with respect to the input quantities are calculated as well as their values. As such 
the bias can be applied using the chain rule as is illustrated schematically for the input above in 
figure~\ref{fig:code-apply}. During this final application step the code runs through the list of currently 
active Actions in reverse as it accumulates the forces on the atoms.  These
forces are ultimately passed back to the MD code and added to the forces from
the interatomic potential.  A special treatment is used
for the {\verb ENERGY } CV \cite{PhysRevLett.104.190601}: instead of explicitly computing the derivatives, PLUMED takes advantage of the fact that the derivative of the potential energy is just minus the force.

PLUMED can also be used to implement multiple replica schemes such as
parallel tempering metadynamics \cite{buss+06jacs}, multiple walkers
metadynamics \cite{rait+06jpcb} and bias exchange metadynamics
\cite{pian-laio07jpcb}.  At present this can 
only be done using the GROMACS engine. However, the implementation is designed to
be of general applicability and is based on the MPI library \cite{mpi}.

\subsection{PLUMED 2 as a set of standalone tools}
\label{sec:tools}

The simplest way to use PLUMED is to download the source from our website, 
compile it and then use it as a standalone tool.  
When PLUMED compiles it generates a single executable called
{\verb plumed }. This minimizes the number of clashes between PLUMED and
other programs. Furthermore, a suffix can be added to the name of the executable
so users can have multiple coexisting versions of PLUMED.

PLUMED's command line tools are run by invoking the {\verb plumed }
executable followed by the name of the tool of interest i.e. using the
following command:
\begin{verbatim}
plumed toolname
\end{verbatim}
A list of available tools can be retrieved using the command:
\begin{verbatim}
plumed help
\end{verbatim}

All the command line tools share a similar syntax and a short help for a
particular command can be
obtained using the command:
\begin{verbatim}
plumed toolname --help
\end{verbatim}

The PLUMED executable includes a simple 
Lennard-Jones MD code ({\verb simplemd }) which can be used to test the
on-the-fly analysis and 
enhanced sampling algorithms, a tool called {\verb driver }  than can be used to
analyze trajectories and a tool called {\verb sum_hills } that should
be used to analyze the results from
metadynamics simulations. 
It is straightforward to add further analysis programs as and when they
are required.
It is important to remember, however, that most of the code's functionality can
be explored using the {\verb driver } option.

\subsection{The interface with the MD code}
\label{sec:code-interface}

One of the principal strengths of PLUMED is the ability to interface the
plug-in 
with a variety of different MD, Monte Carlo and other modeling tools. This
flexibility helps enormously when it comes to disseminating new techniques 
and also allows PLUMED to serve as a platform for cross validating different 
codes and methods.

\begin{figure}
\centering
\includegraphics[height=8cm]{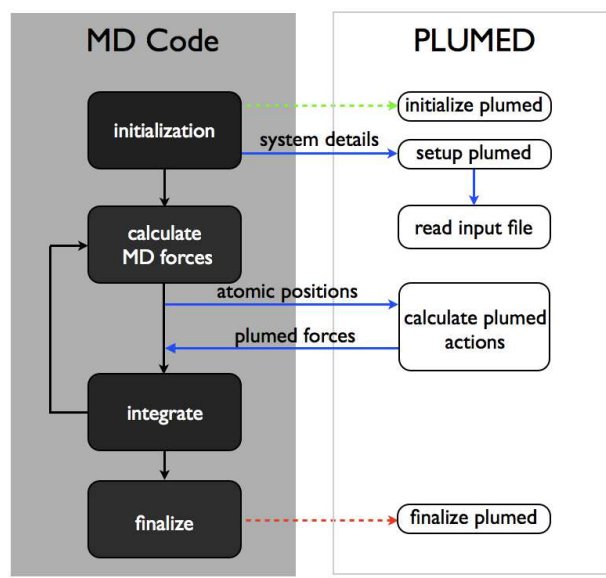}
\caption{A schematic representation of the interface between PLUMED and an MD engine.  
The colors of the arrows indicate the particular PLUMED routine that the MD code should call to perform 
the task.  Green arrows indicate a call to plumed\_create, red arrows indicate a call to plumed\_finalize 
and blue arrows indicate multiple calls to plumed\_cmd. }
\label{fig:MDplumed}
\end{figure}

PLUMED 1 was designed to compile at the same time as the underlying MD engine. This allowed us to 
reuse the underlying MD codes routines for calculating quantities such as
distances, angles and 
torsions and their derivatives. However, it became problematic when implementing CVs and methods 
that were reliant on external libraries as to do this in a way that was
transferable required one to modify 
the makefiles for all the MD codes supported by PLUMED - a laborious and largely
thankless task. A 
consequence of this was that interfaces to the various codes grew at different rates and perhaps 
inevitably a bias was introduced towards those codes used by the majority of the
developers.
 To resolve 
this, we decided that PLUMED 2 would have a single, standard interface, which
is used with all the MD codes and which is not biased towards any
particular package. The PLUMED routines that make up this
interface serve only to communicate data between the MD code and PLUMED. 
This makes it possible to compile PLUMED as set of static object files, as a
dynamic library and even as a standalone post-processing tool. 

In spite of the changes in the way that it is compiled and linked to the MD
engine the points where the MD engine calls PLUMED are the same as in
the previous version.  As shown in figure~\ref{fig:MDplumed} PLUMED is called at
three separate 
points in the MD code. The first is during initialization, the second
occurs just after each force 
evaluation and the third is at the end of the simulation. Three PLUMED routines can be called from within 
the MD engine.  The first of these routines initializes the code and creates the PLUMED object, while the 
last destroys the PLUMED object and frees the memory.  The remaining routine is used in both the ``setup 
plumed'' and ``calculate PLUMED Action'' phases.  It is a generic function that takes a character string 
and a void pointer as arguments.  It is used to pass data between the MD engine
and PLUMED. This 
mode of passing allows us to both retrieve data from the MD code and to modify the MD engine's 
variables as is required for many biased MD methods. The simplicity of this interface makes it easy to 
reuse PLUMED in the many different MD engines used by the community.
Furthermore, there is now some guarantee that the interface will continue
to work when PLUMED is substituted by a different, future version.  Most
importantly, the guidelines for incorporating PLUMED into any MD engine are now
very straightforward.  In fact we have a step-by-step guide in the developer
manual that explains how to incorporate PLUMED into any MD engine.

PLUMED's simplified interface makes it much more straightforward to add the 
option to pass more data between the MD engine and
PLUMED. For instance it would be relatively straightforward to add functionality
to pass the atomic velocities from the MD engine to 
PLUMED. This opens the door to using PLUMED for many other methodologies 
and to a massive extension in the range of functionalities provided by PLUMED.

\section{Examples}
\label{sec:examples}

\subsection{Steered MD on a system at the DFT level of theory}

At heart PLUMED 2, like its predecessor, is a code for doing enhanced sampling calculations.  Hence, 
in this first example we showcase these functionalities by demonstrating how we
can use the plug-in to 
examine the SN2 reaction between a methyl-chloride molecule and a chlorine atom.  The potential in 
this system was calculated using the BLYP density functional as implemented in
the PW code from Quantum ESPRESSO 5.0 \cite{QE-2009}.  Obviously, the
potential energy of the SN2 transition state is 
very high so we are unlikely to see the reaction in an unbiased MD simulation.
To resolve this we 
used steered MD to force the reaction to occur.  This method works by applying a harmonic potential, the 
equilibrium position of which moves at a constant velocity.  In our SN2 reaction
example this potential 
is a function of the distance between the carbon atom and one of the two chlorine atoms and as the 
simulation progresses its equilibrium position moves along this coordinate.  As the system is attached to 
this moving potential its motion obviously forces the system to change the length of the carbon-chlorine 
bond, which in turn makes the reaction occur.  Clearly, the reaction only occurs because the potential 
does some work on the system.  It is easy to calculate how much work the potential performs on the 
system using:

\begin{equation}
W(s,t) = \int_0^t \textrm{d}t' k \left( s(t) - s_0-\frac{s_1 -  s_0}{t_1-t_0}t  \right),
\label{eqn:work}  
\end{equation}
where $s(t)$ is the value of the CV at time $t$,  $s_0$ is the equilibrium
position for the harmonic restraint  
at time $t_0$ and  $s_1$ is the equilibrium position for the harmonic restraint
at time $t_1$. Free energies can be estimated from work values calculated
using equation~\ref{eqn:work} by using the Jarzynski
equality \cite{Jarzynski:1997uj} but this is
beyond the scope of this paper.

The input below instructs PLUMED to do a steered MD calculation using the {\verb MOVINGRESTRAINT }
Action.  

\begin{verbatim}
d12: DISTANCE ATOMS=1,2
d23: DISTANCE ATOMS=2,3
# moving restraint
MOVINGRESTRAINT ...
 ARG=d12
 STEP0=0    AT0=0.31  KAPPA0=200000.0
 STEP1=5000 AT1=0.18
 LABEL=steer
... MOVINGRESTRAINT
PRINT ... 
 FILE=COLVAR ARG=d12,d23,steer.d12_cntr,steer.d12_work 
 STRIDE=1
... PRINT
ENDPLUMED
\end{verbatim}

\begin{figure}
\centering
\includegraphics[height=9cm,angle=0]{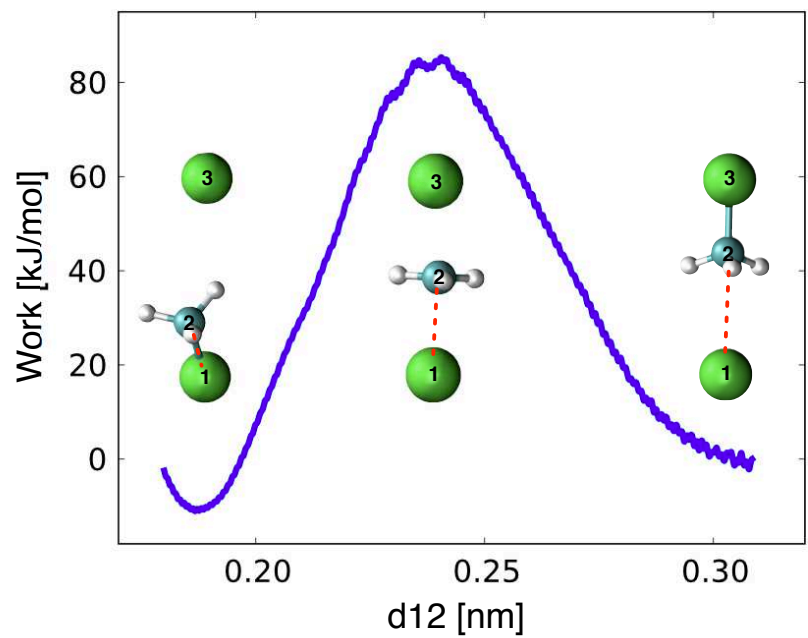}
\caption{Worked performed during a steered MD simulation 
using as CV the distance (red dotted line) between a chlorine atom (atom 1) and the 
carbon atom of the methyl group (atom 2).}
\label{fig:steer-fig}
\end{figure}

This input instructs PLUMED to calculate two distances but only one (the distance between atoms 
1 and 2: {\verb d12 }) is used in the moving restraint.  Initially, the 
equilibrium position for the harmonic potential is at {\verb d12 } equal to 0.31~nm,
thus ensuring that atoms 1 and 2 are not bonded. 
The center of the potential then moves at constant
velocity to a value for {\verb d12 } of 0.18~nm during the course of 
the 5~ps (5000 step) simulation.  The potential is a harmonic spring with a
force constant of 2$\times 10^{5}$~kJ~mol$^{-1}$~nm$^{-2}$ so the motion of the center
causes the system to move towards configurations in which the bond is formed.  
During the simulation PLUMED  generates a file called {\verb COLVAR }.  This file is 
produced thanks to the {\verb PRINT } Action, which should always be used to output quantities calculated by the 
various Actions in the input file.  The form of the {\verb PRINT } command in the input file above ensures that the 
values of the distances {\verb d12 } and {\verb d23 } are output at every step  together with the equilibrium position for the 
moving harmonic potential {\verb steer.d12_cntr } and the work {\verb steer.d12_work }, which is calculated using 
equation~\ref{eqn:work}.  Figure~\ref{fig:steer-fig} shows a plot of the work performed by the potential as 
it moves along the {\verb d12 } coordinate (from right to left).
As one would expect for a SN2 reaction of this kind a maximum for the work 
appears when the methyl group is equidistant from the two chlorine atoms.

\subsection{Using functions of collective variables}

{\verb MATHEVAL } is one of the most powerful Actions in PLUMED 2. This Action can be used to construct 
linear and non-linear functions of CVs with the libmatheval library \cite{libmatheval}. 
The example input below shows how complex CVs can be calculated without 
making any modifications to PLUMED by using this Action in tandem with the simple CVs 
that are already present:

\begin{verbatim}
# just declare the RMSD^2 for five structures
t1: RMSD REFERENCE=c_1.pdb TYPE=OPTIMAL SQUARED
t2: RMSD REFERENCE=c_2.pdb TYPE=OPTIMAL SQUARED 
t3: RMSD REFERENCE=c_3.pdb TYPE=OPTIMAL SQUARED
t4: RMSD REFERENCE=c_4.pdb TYPE=OPTIMAL SQUARED
t5: RMSD REFERENCE=c_5.pdb TYPE=OPTIMAL SQUARED
# calculate the sum of the exponential of the five RMSDs
MATHEVAL ...
 LABEL=dwn 
 ARG=t1,t2,t3,t4,t5
 VAR=d1,d2,d3,d4,d5
 FUNC=(exp(-770*d1)+exp(-770*d2)+exp(-770*d3)+exp(-770*d4)+exp(-770*d5))        
 PERIODIC=NO
... MATHEVAL
# now calculate the indexed sum
MATHEVAL ...
 LABEL=up 
 ARG=t1,t2,t3,t4,t5
 VAR=d1,d2,d3,d4,d5
 FUNC=(exp(-770*d1)+2*exp(-770*d2)+3*exp(-770*d3)+4*exp(-770*d4)+5*exp(-770*d5))        
 PERIODIC=NO
... MATHEVAL
# combine them into a progress function
MATHEVAL ...
 LABEL=s ARG=up,dwn VAR=u,d PERIODIC=NO
 FUNC=u/d        
... MATHEVAL
# combine them into the distance from the path
MATHEVAL ...
 LABEL=z ARG=dwn VAR=d  PERIODIC=NO
 FUNC=-(1./770.)*log(d)        
... MATHEVAL
# some printout
PRINT ARG=s,z STRIDE=100 FILE=colvar FMT=%
# do metadynamics with welltempered and adaptive hills
METAD ...
 HEIGHT=1.2 SIGMA=0.02 PACE=60 
 ARG=s,z ADAPTIVE=GEOM BIASFACTOR=5 TEMP=300
... METAD
\end{verbatim}

This input file uses {\verb MATHEVAL } multiple times in order to generate non-linear combinations of the mean 
square displacements from a number of different reference points.  The particular non-linear 
combination we are creating are the path CVs of 
Branduardi {\it et al.} \cite{bran+07jcp}.  These variables measure the progress $s$ along some curvilinear path in the 
high-dimensional space:  
\begin{equation}
s = \frac{\sum_{i=1}^5 i \exp(-\lambda M_i({\bf X}))}
             {\sum_{i=1}^5  \exp(-\lambda M_i({\bf X}))},
\end{equation}
and the distance $z$ from the path:
\begin{equation}
z = -\frac{1}{\lambda}\ln {\sum_{i=1}^5 \exp(-\lambda M_i({\bf X}))},
\end{equation}
where $M_i({\bf X})$ is the mean-square deviation, after optimal alignment, of a subset of the atoms 
from a reference structure denoted by the $i$ index. The parameter
$\lambda$ is a 
smoothing parameter that can be set by examining the average
distance between adjacent images.
In the example input above, these $M_i$ values are calculated by the Actions
labeled {\verb c1 }, {\verb c2 }, 
{\verb c3 }, {\verb c4 } and {\verb c5 }.

To be clear, path CVs are now quite widely used by the community and PLUMED 2 
contains a simpler command ({\verb PATHMSD }) that can be used to calculate these CVs.  Consequently, much 
of the complexity in the input file above can be avoided by the casual user.  However, we believe this 
example is still instructive as it demonstrates how complicated CVs can be generated in 
the input file and how PLUMED and the matheval library automatically look after the derivatives of these 
functions.  {\verb MATHEVAL } allows one to try new CV combinations without modifying the 
interior of the code.  

\begin{figure}
\centering
\includegraphics[height=9cm,angle=0]{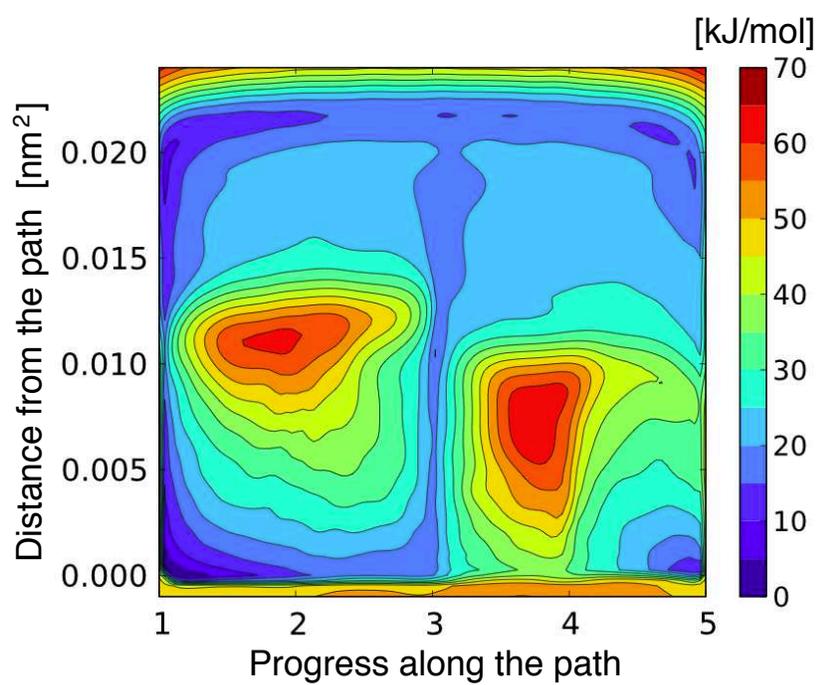}
\caption{The FES of alanine dipeptide as a function of the path CVs 
discussed in the text.  This FES was calculated using a metadynamics simulation with 
adaptive hills.}
\label{fig:diala-fes}
\end{figure}

The path defined in the above input is an optimized path for alanine dipeptide in vacuum that connects 
the metastable states C7eq and Cax. We modeled this system using the CHARMM27 forcefield 
\cite{charmm27a,charmm27b} as implemented in GROMACS 4.5.5 \cite{gromacs}.  In order to see transitions between the $\rm C7_{eq}$ and $\rm C_{ax}$ states we 
performed a 3~ns well-tempered metadynamics simulation \cite{bard+08prl}
using the geometry-adapted Gaussians 
scheme introduced by Branduardi {\it et al.} \cite{adapt-gau}.  The free-energy surface (FES) shown in 
figure~\ref{fig:diala-fes} was extracted by post-processing this simulation
using a Torrie-Valleau 
correction \cite{torri-valle77jcp,adapt-gau} and can be easily calculated using the post processing tools in PLUMED 2. 

\subsection{Using a function of a distribution of CVs}

\begin{figure}
\centering
\includegraphics[height=9cm]{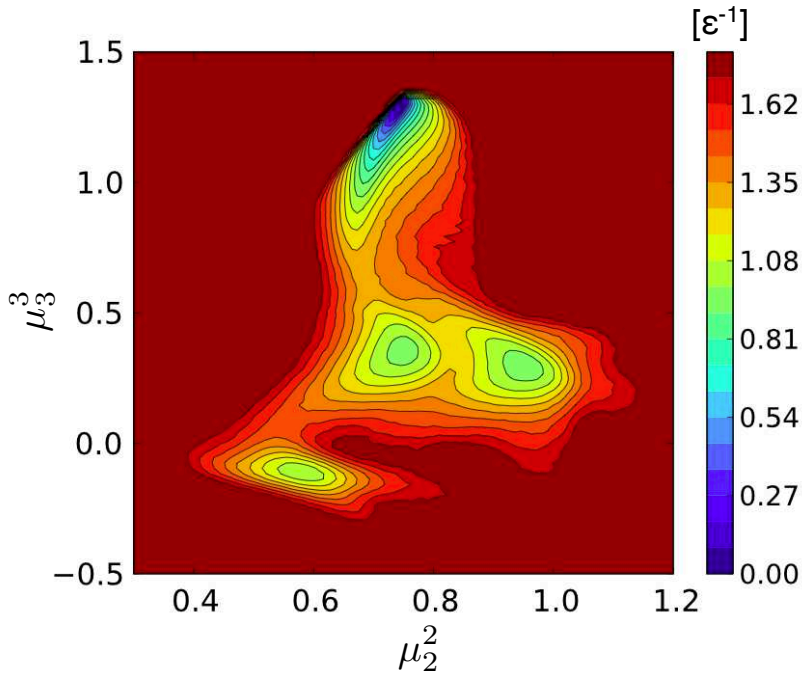}
\caption{The FES for a seven-atom Lennard-Jones cluster as a function of the second and third moments of 
the distribution of coordination numbers. These CVs map the four known minima for this structure to 
different parts of the 2D plane. This FES was obtained by reweighting the result of a 
higher temperature MD simulation.}
\label{fig:lj7-fes}
\end{figure}

When devising CVs for clusters or bulk materials
it is important to
remember that the value of the CV should not change when the labels on atoms of 
the same type are exchanged.  If a process such as nucleation is examined
without considering this symmetry a large number of pathways connecting the
liquid to the crystalline state will be found as the critical nuclei can be
formed from any combination of the atoms in the system.  In fact the number of
pathways from the liquid to solid state is not analytically countable {\it a
priori}.  To examine these sorts of problems we therefore need global CVs that
measure how a distribution of atomic/molecular order parameters changes.
A common strategy is to calculate the values of some 
symmetry function ({\it e.g.}~coordination numbers, Steinhardt parameters \cite{steinhardt}) for each atom and to use the 
average of the distribution as a global CV \cite{quig+08jcp}. It is known, however, that this often forces reactions to take 
place via a highly-concerted, unphysical mechanism. Furthermore, the mean will not necessarily 
separate all the structurally distinct configurations of a cluster or bulk solid. The average is not the only 
quantity one can calculate from a distribution of symmetry functions and there is strong evidence that 
calculating functions other than the average allows you to examine interesting phenomena \cite{rdf1,lj-recon,recent-smap}.  Hence, 
within PLUMED 2 we provide tools to calculate the average of a distribution of symmetry functions, the 
moments of the distribution, the number of symmetry functions less than a certain value and so on all 
within the MultiColvar class. To demonstrate how this class can be used in practice we show here an 
example calculation on a well studied, two-dimensional, seven-atom Lennard-Jones cluster.  This cluster 
is known to have four minima that appear at distinct points on the two dimensional plane described by 
the second ($\mu_2^2 = \frac{1}{N} \sum_{i=1}^N \langle c_i - \overline{c}\rangle^2$) and third 
($\mu_3^3 = \frac{1}{N} \sum_{i=1}^N \langle c_i - \overline{c}\rangle^3$) moments of the distribution of 
coordination numbers \cite{lj-minima,recon}. These quantities can be calculated by PLUMED 2 using the following command:
\begin{verbatim}
COORDINATIONNUMBER ...
 SPECIES=1-7 
 SWITCH={RATIONAL R_0=1.5 NN=8 MM=16}
 MOMENTS=2-3
 LABEL=c1
... COORDINATIONNUMBER  
\end{verbatim}
Coordination numbers for each of the seven atoms are calculated using:
\begin{equation}
c_i(\mathbf{X}) = \sum_{j \ne i} \frac{ 1 -
\left(\frac{d_{ij}(\mathbf{X})}{1.5}\right)^8 }{ 1  -
\left(\frac{d_{ij}(\mathbf{X})}{1.5}\right)^{16} },
\end{equation}
where $d_{ij}(\mathbf{X})$ is the distance between the $i$-th atom and the
$j$-th atom.
Then, once the coordination numbers have been calculated, the moments (and their derivatives) are 
calculated using the equations in the paragraph above and are stored so that they can be referenced in 
other Actions as {\verb c1.moment_2 } and {\verb c1.moment_3 }. 

To calculate the FES of seven-atom Lennard-Jones cluster as a function of the moments we ran an MD 
simulation at 0.2~$\frac{k_B T}{\epsilon}$.  The final result of the calculation is shown in 
figure~\ref{fig:lj7-fes}.  The PLUMED 2 input for this calculation was as follows:

\begin{verbatim}
#
# instructs plumed to use LJ units for length and energy
#
UNITS NATURAL
#
# boundaries to prevent the system from subliming
#
COM ATOMS=1-7 LABEL=com
DISTANCE ATOMS=1,com LABEL=d1
UPPER_WALLS ARG=d1 AT=2.0 KAPPA=100.
DISTANCE ATOMS=2,com LABEL=d2
UPPER_WALLS ARG=d2 AT=2.0 KAPPA=100.
DISTANCE ATOMS=3,com LABEL=d3
UPPER_WALLS ARG=d3 AT=2.0 KAPPA=100.
DISTANCE ATOMS=4,com LABEL=d4
UPPER_WALLS ARG=d4 AT=2.0 KAPPA=100.
DISTANCE ATOMS=5,com LABEL=d5
UPPER_WALLS ARG=d5 AT=2.0 KAPPA=100.
DISTANCE ATOMS=6,com LABEL=d6
UPPER_WALLS ARG=d6 AT=2.0 KAPPA=100.
DISTANCE ATOMS=7,com LABEL=d7
UPPER_WALLS ARG=d7 AT=2.0 KAPPA=100.
#
# calculates moments of the coordination number distribution
#
COORDINATIONNUMBER ...
 SPECIES=1-7
 MOMENTS=2-3 
 SWITCH={RATIONAL R_0=1.5 NN=8 MM=16}
 LABEL=c1
... COORDINATIONNUMBER 
#
# calculate histograms from the moments
#
HISTOGRAM ...
 ARG=c1.moment_2,c1.moment_3 STRIDE=10
 REWEIGHT_TEMP=0.1 TEMP=0.2
 GRID_MIN=0.2,-0.5 GRID_MAX=1.2,1.7 GRID_BIN=200,440
 BANDWIDTH=0.01,0.01 KERNEL=triangular
 GRID_WSTRIDE=10000000 GRID_WFILE=histo
... HISTOGRAM
\end{verbatim}

For this calculation we worked in Lennard-Jones units so lengths are in units of $\sigma$, energies are 
in units of $\epsilon$ and times are in units of $t^*=\sqrt{\frac{m \sigma^2}{\epsilon} }$.  We introduced a set 
of restraints on the distances between the positions of the atoms in the cluster and the center of mass in 
order to prevent the cluster from subliming. We then ran $10^7$ steps of MD at a temperature of 
0.2~$\frac{k_B T}{\epsilon}$ using the Lennard-Jones MD code ({\verb simplemd }) that forms part of PLUMED. The temperature in these simulations was 
kept fixed using a Langevin thermostat with a relaxation time of 1.0~$t^*$ and the timestep was 
0.005~$t^*$.  We extracted a FES at the lower temperature of 0.1~$\frac{k_B T}{\epsilon}$ 
by reweighting the probability distribution calculated at the higher
temperature, $P(\mathbf{s})$, using 
$\exp(-( \frac{E(\mathbf{X})}{0.1} - \frac{E(\mathbf{X})}{0.2})
)P(\mathbf{s})$. This analysis is done on the fly using 
PLUMED 2's histogram utility, which performs kernel density estimation with triangular kernel functions.  
The FES shown in figure~\ref{fig:lj7-fes} is an average from 16 such runs.  Each of these 
calculations was started from a different, equilibrated configuration. The standard deviations between the 
estimates of the free energies obtained from the individual runs were on the order of 0.05~$\epsilon$ over 
the whole FES. 

\section{Adding a new functionality to PLUMED 2}
\label{sec:adding}

As shown in the examples above, the fact that the PLUMED 2 input file is written in a pseudo scripting 
language allows users to use a wide variety of CVs and methods. Even so, some users will eventually 
find themselves requiring something that is not implemented and that requires some coding.  It is 
important to note at the outset that the object-oriented style of the code makes it straightforward to reuse 
features written by others. Developers should thus look carefully at the developer manual before starting 
programming so as to work out what functionality from the core can be reused.  In particular, if they are 
interested in implementing a CV, a function of a CV, a biasing method or a method for analyzing the 
trajectory the developer manual contains step by step instructions as to how to go about performing these tasks.

The creation of a new piece of functionality usually involves the creation of 
a single new source code file, 
which, in the majority of cases, provides a definition for a new Action object. This file should eventually 
contain the definition of the class, the definitions for the methods in the class and the documentation for 
the new piece of functionality. Furthermore, any new Action object should inherit either directly or 
indirectly from the Action class. In this way new functionality can be added to PLUMED 2 through 
dynamic polymorphism and hence without modification of the PLUMED core. 
Methods in the new class with certain special name, {\it e.g.}~{\verb calculate }, 
will then be called at certain key points during the execution of each MD step
and the new Action will be properly integrated into PLUMED. In addition,
in classes that inherit from Action protected routines can be reused to output
to the log file, to parse the input and to detect errors.

From discussions with the users we found that one reason for PLUMED 1's success was the provision of 
a detailed manual.  We thus felt it was important to ensure that good documentation was provided.  Each 
Action in PLUMED 2 is described in the manual in a single page of HTML, which gives a brief description 
of what the Action calculates and what it can be used for, a description of the list of keywords for the 
Action and finally some examples of the Action's use.  These HTML pages are generated from 
descriptions in the individual source code files using 
Doxygen \cite{doxygen}.  We felt that keeping the user 
documentation and source code together in this way was a good way of
encouraging people to keep the manual consistent with the code. To write the
documentation developers have to write a short paragraph of text and to 
provide some examples before the executable code in the source code file.  In writing this text 
developers make use of Doxygen, which provides simple \LaTeX -like commands that allow one to 
include equations and references. The list of available options or keywords for each Action and 
descriptions of the keywords is required to be placed inside the source as this information is also used to 
provide useful error messages.  In addition, T. Giorgino has reused these keyword descriptions to 
provide templates for PLUMED GUI \cite{toniG}.  Finally, a side benefit of
having keyword descriptions in the 
source is that a single piece of documentation for an often used keyword can be used in the descriptions of 
multiple Actions. 

Developers modifying the code are strongly encouraged to use regtests.  There is now a simple 
procedure for including a new test and all tests are run using a single script. Regularly running regtests 
as the code is developed  allows one to ensure that new features are not breaking established 
functionalities in the code.  Even developers working on code that is not being
shared will benefit if they have regtests for new functionalities as they can
use regtests to 
ensure that their new functionality continues to work when code is updated upstream.  

Lastly, if users have exotic CVs or methods that they have used PLUMED to calculate we 
encourage them to share their experiences with the community.  To help in this we have provided 
functionality for users and developers to share information through tutorials that can be added 
to the manual for the code. 

\section{Conclusion and outlook}

New simulation techniques are oftentimes not rapidly exploited in the wider community because easy-to-use 
implementations of them are not readily available. PLUMED thus performs a vital service by disseminating 
new methods to the widest possible community of users in a form that can be added to many of the 
available MD engines.

With PLUMED 2 we have removed many of the limitations that were in PLUMED 1. 
It is now far easier to use new CVs as complex combinations of
variables can be assembled directly from the input file. In addition, the
performance of the code has been improved by parallelizing variables and by
calculating them only when needed.
One further, particularly-important improvement is that the
interface between PLUMED and the MD engines has been simplified considerably.  
This will make maintenance easier and, given that PLUMED can now be compiled as a dynamic library, 
even give MD engine developers the option of providing their codes with the PLUMED interface in place 
already. This more flexible interface also makes it more straightforward to
change the amount of 
information passed between the MD code and PLUMED.  Hence, PLUMED could now easily be 
extended and used to implement features such as thermostats
or force-fields. Furthermore, we believe that the internal flexibility of the new version
will allow many more scientists to contribute new CVs,
free-energy methods, and analysis tools.  This will thus further foster the
development of new techniques for MD. 

\section{Availability}

PLUMED 2 can be downloaded from 
\href{http://www.plumed-code.org}{www.plumed-code.org}.
To best serve the community of PLUMED users and developers we have slightly changed the way the 
code is delivered.  Now, as well as having stable releases of the code that can be downloaded from the 
website, we also provide read-only access to a git repository containing the development version of the 
code.  For this reason we now have two mailgroups: the old 
\href{mailto:plumed-users@googlegroups.com}{plumed-users@googlegroups.com} 
and 
\href{mailto:plumed2-git@googlegroups.com}{plumed2-git@googlegroups.com}.  
The second is designed for users and developers experimenting with the latest development version of 
the code which is available from \href{http://www.plumed-code.org/git}{www.plumed-code.org/git}.  Users 
who wish to distribute their own additional functionality for PLUMED can do so by either contacting the 
developers or by providing the additional code through their own websites.    

\section{Acknowledgments}

The reworking of PLUMED 2 that has been described in this paper was made possible by a CeCAM 
grant and by a ``Young SISSA Scientists' Research Projects'' scheme 2011-2012 that
was promoted by 
the International School for Advanced Studies (SISSA), Trieste, Italy. Using funding from these sources 
we were able to organize a PLUMED tutorial, a developer meeting and a user meeting. These forums 
provided the opportunity to meet the various users and developers of the code and to find out the various 
problems with the original package.  We would like to acknowledge all the people who attended these 
meetings for their feedback as well as the many users who have provided valued contributions to our 
mailing list. In particular, we would like to thank Paolo Elvati, Toni
Giorgino, Alessandro Laio, Layla Martin-Samos, Michele Parrinello, Fabio
Pietrucci and Paolo Raiteri
for making
time to discuss the code privately with the 
authors.  In addition, GB and DB acknowledge the HPC-EUROPA2 project no. 228398, GB 
acknowledges MIUR grant FIRB (Futuro in Ricerca no. RBFR102PY5)
and the European Research Council (Starting Grant S-RNA-S, no. 306662), and CC
acknowledges support from a 
Marie Curie Intra European Fellowship.

\end{document}